\documentclass[%
 reprint,
 amsmath,amssymb,
 aps,
]{revtex4-2}

\usepackage{graphicx}
\usepackage{dcolumn}
\usepackage{bm}

\begin{document}

\preprint{APS/123-QED}

\title{Light-Matter Interaction Near the Schwinger Limit Using Tightly Focused Doppler-Boosted Lasers}

\author{Neïl Zaïm$^1$}
\author{Antonin Sainte-Marie$^1$}
\author{Luca Fedeli$^1$}
\author{Pierre Bartoli$^1$}
\author{Axel Huebl$^2$}
\author{Jean-Luc Vay$^2$}
\author{Henri Vincenti$^1$}
\affiliation{$^1$Université Paris-Saclay, CEA, CNRS, LIDYL, 91191 Gif-sur-Yvette, France}
\affiliation{$^2$Lawrence Berkeley National Laboratory, Berkeley, CA 94720, United States of America}

\date{\today}

\begin{abstract}
The Schwinger limit could be approached by focusing to its diffraction limit the light reflected by a plasma mirror irradiated by a multi-petawatt laser. We explore numerically the interaction between such intense light and matter. We find that the interaction with a relativistic counterpropagative electron beam would enable the exploration of the fully nonperturbative regime of strong-field quantum electrodynamics (SF-QED), while the interaction with an initially solid target leads to a profusion of SF-QED effects that retroact on the laser-plasma interaction. We observe in both scenarios the formation of relativistic attosecond electron-positron jets with very high densities.
\end{abstract}

\maketitle

Multi-petawatt (PW) femtosecond (fs) lasers~\cite{dans19} can nowadays reach intensities exceeding 10$^{23}$~W/cm$^2$~\cite{yoon21}. The interaction of such high-intensity lasers with matter can no longer be described purely by classical electrodynamics, enabling unexplored interaction regimes in which strong-field quantum electrodynamics (SF-QED) phenomena must be considered~\cite{dipi12,gono22,fedo22}.

The dominant SF-QED processes are the discretized emission of high-energy photons by electrons and positrons (nonlinear inverse Compton scattering~\cite{erbe66}) and the decay of high-energy photons into electron-positron pairs (nonlinear Breit-Wheeler~\cite{erbe66,brei34}) in the presence of strong fields. The importance of these processes is determined by the ratio $\chi$ between the electric field $E_{rest}$ in the rest frame of a lepton (electron or positron) and the critical field of SF-QED - the so-called Schwinger field~\cite{schw51} $E_S \approx 1.32 \times 10^{18}$~V/m. Strong-field effects become substantial whenever this quantum parameter $\chi = E_{rest}/E_S$ exceeds unity. In the coming years, it is expected that laser-electron beam collisions - a configuration that maximizes the $\chi$ parameter for a given laser - will allow the detailed study of these basic processes of SF-QED~\cite{lobe17,cole18,pode18,abra19,yaki19}.

Going further, some of the most attractive prospects of the study of SF-QED are (i) the experimental access to its fully nonperturbative regime, conjectured to occur when $\chi > 1600$~\cite{fedo17}, and for which there is no universally accepted theory~\cite{ilde19b,miro20}, (ii) the exploration and understanding of so-called QED-plasma states~\cite{zhan20}, defined by the self-consistent interaction between SF-QED processes and plasma phenomena, which is relevant for understanding plasma states close to compact astrophysical objects (such as black holes, magnetars or neutron stars) and their electromagnetic signatures~\cite{mesz01,uzde14,chen23}, and (iii) the possibility to develop novel ultra-short and bright sources of antimatter and high-energy photons with applications to high energy physics or nuclear science~\cite{vran19,gu19}. Unfortunately, these prospects are projected to remain out of the reach of the schemes commonly considered for probing SF-QED (laser-electron collisions~\cite{lobe17,qu21}, seeded laser-laser collisions~\cite{neru11,gris16,zhu16} and laser-solid interactions~\cite{ridg12}) until the advent of lasers with peak powers exceeding several tens of PW~\cite{zhan20}. For this reason, alternative approaches aimed at making the most of existing laser powers and energies are being actively explored~\cite{gono13,gono17}.

\begin{figure*}[t]
\includegraphics[width=2.\columnwidth]{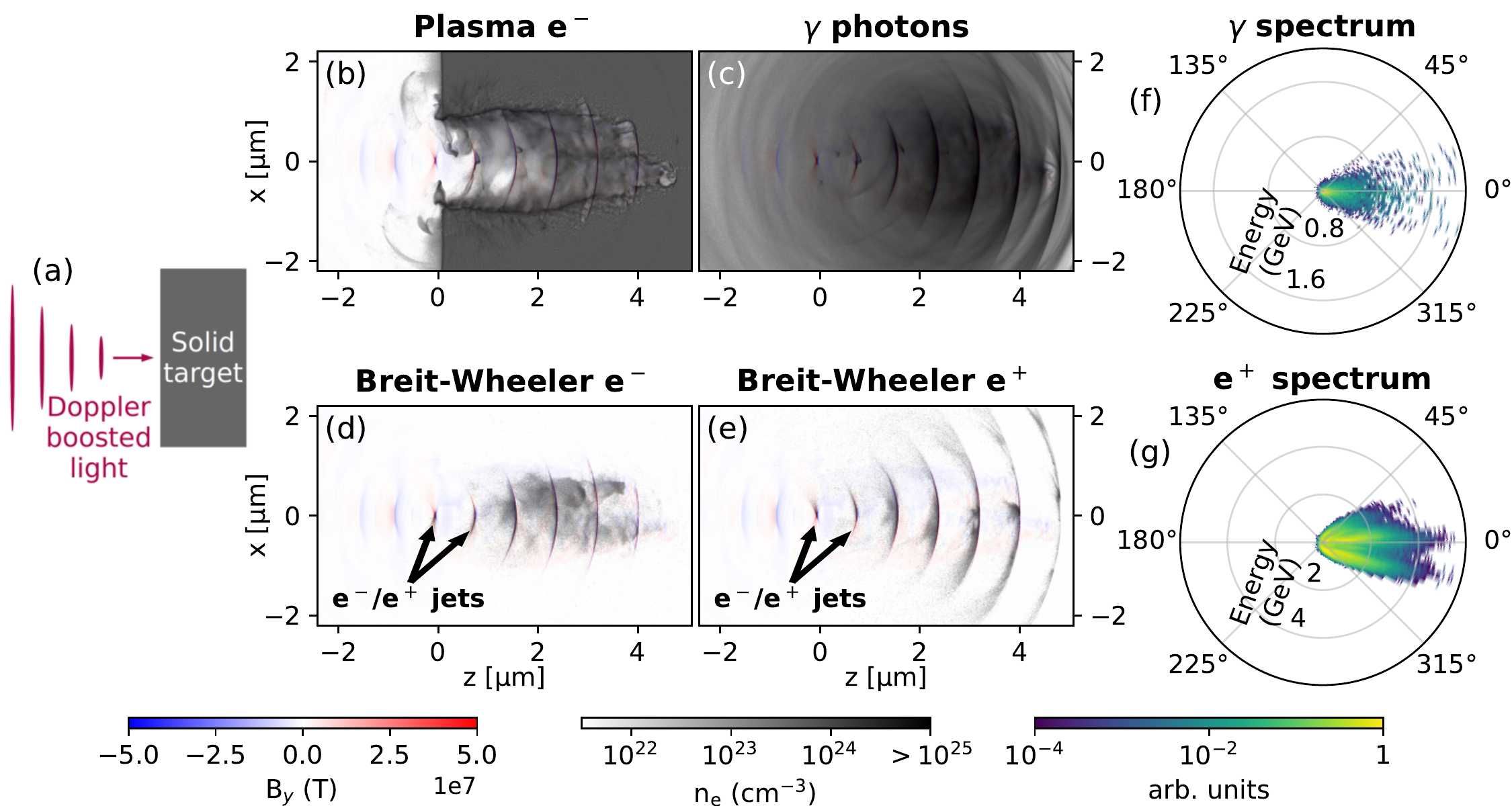}
\caption{Simulation of the interaction between a boosted laser with 1.2$\times$10$^{28}$~W/cm$^2$ peak intensity and a solid target. (a) Schematic illustration of the interaction. (b-e) Snapshots showing the boosted laser magnetic field combined with (b) the initial plasma electron, (c) the $\gamma$ photon, (d) the Breit-Wheeler electron or (e) the Breit-Wheeler positron density during the interaction. (f-g) $\gamma$ photon (f) and positron (g) angularly resolved spectrum at the end of the simulation.}
\label{FigSolid}
\end{figure*}

Among these, the possibility of boosting the intensity of a PW laser by Doppler effect on a plasma mirror has recently received significant attention~\cite{gord05,drom09,gono11,vinc19,quer21,fede21b,sain22,mark23}. This boosting technique relies on the interaction between a PW laser and a flat solid target that is ionized into a plasma mirror. During this interaction, two processes result in a strong intensification of the reflected laser. First, the plasma mirror surface oscillates non-linearly with the same periodicity as the incident laser field and with relativistic velocities. Thus, during part of this cycle, the incident laser is effectively reflected by a relativistically moving mirror, which increases its frequency by Doppler effect~\cite{teub09,thau10}. In the time domain, this corresponds to a temporal compression of the laser energy into a train of attosecond pulses. Second, during the interaction the plasma mirror is carved into a parabolic shape by the transversely inhomogeneous laser radiation pressure. This plasma surface curvature leads to a focusing of the reflected laser in a much smaller volume than possible with the initial laser frequency.

Massively parallel Particle-In-Cell (PIC)~\cite{bird04} simulations have recently demonstrated that this compression of the laser energy is associated with intensity gains that can exceed 3 orders of magnitude~\cite{vinc19}. Even more recently, it was shown that the newly accessible intensity range between 10$^{24}$ and 10$^{26}$~W/cm$^2$ could be leveraged to strongly increase the signatures of basic SF-QED processes in laser-solid interactions~\cite{fede21b}.

Yet, one of the most appealing perspectives opened by the Doppler boosting technique is the practical possibility to attain intensities between 10$^{27}$~W/cm$^2$ and 10$^{29}$~W/cm$^2$ with already existing laser powers~\cite{quer21}. Reaching such high intensities requires that the boosted lasers be focused close to their diffraction limit, which is a challenge given the small spatial scales involved ($\approx$ 10-100~nm). Nevertheless, schemes are currently being explored to focus the lasers reflected from a plasma mirror as tightly as possible~\cite{chop21}, either by tuning the laser-plasma interaction or by using external XUV focusing optics. Additionally, Doppler boosting appears today as the only realistic path for obtaining electric fields close to the Schwinger critical field (which corresponds to an intensity of 4.65$\times$10$^{29}$~W/cm$^2$) \textit{in the laboratory frame}.

In this context, the present letter aims at understanding the basic physics of Doppler-boosted laser-matter interaction at intensities close to the Schwinger limit. In the following, we present a general overview of the interaction scenarios that can be envisaged with an aberration-free Doppler boosted laser focused close to its diffraction limit on matter at intensities between 10$^{27}$~W/cm$^2$ and 10$^{29}$~W/cm$^2$. Such an idealization is important as a first approach to: (i) understand the main physical regimes that come at play in these scenarios and their associated signatures and (ii) help define and motivate future research directions with Doppler-boosted lasers.

\begin{figure*}[t]
\includegraphics[width=2.\columnwidth]{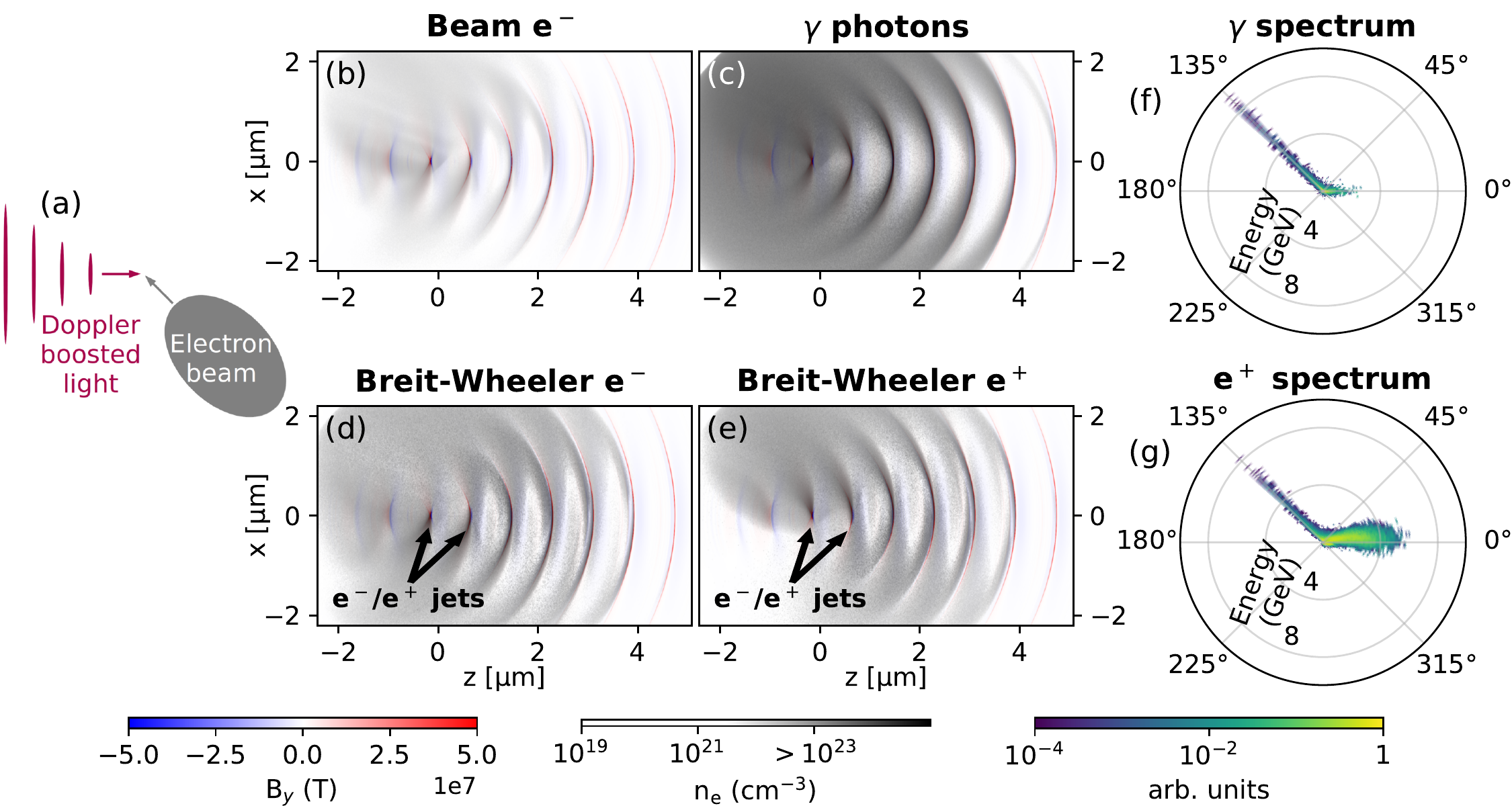}
\caption{Simulation of the interaction between a boosted laser with 1.2$\times$10$^{28}$~W/cm$^2$ peak intensity and a counterpropagative 10~GeV electron beam. (a) Schematic illustration of the interaction. (b-e) Snapshots showing the boosted laser magnetic field combined with (b) the beam electron, (c) the $\gamma$ photon, (d) the Breit-Wheeler electron or (e) the Breit-Wheeler positron density during the interaction. (f-g) $\gamma$ photon (f) and positron (g) angularly resolved spectrum at the end of the simulation.}
\label{FigBeam}
\end{figure*}

To this end, we have performed 2D PIC simulations, using the code WarpX~\cite{myer21,fede22} coupled to the PICSAR-QED library~\cite{fede22b}, of the two most common \textit{single-laser} scenarios that are envisioned to probe SF-QED: the interaction of a boosted laser with a solid target, illustrated in Fig.~\ref{FigSolid}(a) and the collision of a boosted laser with a relativistic electron beam, illustrated in Fig.~\ref{FigBeam}(a). The standard QED-PIC algorithm~\cite{gono15} is used in the simulations, with the addition of a particle thinning algorithm~\cite{mura21} to deal with the copious particle creation occurring at such intensities.

We use simple interaction geometries in the simulations. The interaction with a solid, which can be in practice implemented using an L-shaped target~\cite{fede21b}, is performed at normal incidence with a fully ionized SiO$_2$ target of constant density 6.5$\times$10$^{23}$~cm$^{-3}$. The interaction with an electron beam is performed using beam parameters that are expected to soon become available in compact state-of-the-art laser wakefield acceleration facilities~\cite{albe21}: an electron energy of 10~GeV and a total charge of 10~pC spread in a transverse size of 2~$\mu$m and a longitudinal size of 3~$\mu$m. A 45$^{\circ}$ angle is used between the laser and the electrons with the aim of reducing radiative losses in the first attosecond pulses of the boosted laser so that the electron beam can interact with its full energy at the point where the laser reaches its maximum intensity~\cite{blac19}.

Finally, we use the same procedure as in Ref.~\cite{sain22} to model an ideal Doppler boosted laser: its spectrum is obtained from optimized PIC simulations and its spatio-temporal profile is obtained by assuming that each frequency component has a Gaussian transverse shape and is focused to a diffraction-limited spot-size. With the chosen spectrum, wavelengths ranging between 800~nm and 8~nm significantly contribute to the peak intensity. The resulting most intense attosecond pulse has, at focus, a duration of 21 attoseconds and a transverse size of 39~nm in FWHM of the field. We have considered in this work peak intensities between 1.3$\times$10$^{27}$~W/cm$^2$ and 1.2$\times$10$^{29}$~W/cm$^2$, which would correspond to primary laser powers ranging between 0.4~PW and 17~PW according to estimates of the intensity gain that can be achieved with the Doppler boosting technique in the diffraction limited regime~\cite{quer21}. It should be kept in mind however that a more detailed study still needs to be performed to precisely establish the maximum boosted laser intensity that can theoretically be obtained for a given initial laser power. More details regarding the simulation parameters are given in the Supplemental Material~\cite{suppmat}.

Figure~\ref{FigSolid} summarizes the main features of the simulation of the interaction between a 1.2$\times$10$^{28}$~W/cm$^2$ boosted laser and a solid target. We first note that the plasma is underdense for the high-frequency components of the laser. Combined with the effect of relativistic transparency~\cite{mour06,pala12}, this means that a significant fraction of the laser propagates inside the solid, digging a plasma channel nearly void of electrons. The strong fields associated with both the laser and the plasma channel lead to copious $\gamma$ photon and electron-positron pair generation up to $\sim$4~$\mu$m inside the solid, at which point the laser is almost completely absorbed.

The field of the laser is strong enough to directly trap and accelerate along the laser propagation direction~\cite{thev15} the generated pairs to energies up to 6~GeV (see Fig.~\ref{FigSolid}(g)). The Breit-Wheeler electrons and positrons therefore find themselves bunched into very small volumes (approximately 10~nm long and 200~nm wide at the focus of the laser) by the Doppler-boosted laser, leading to the formation of a train of extremely dense (exceeding 10$^{25}$~cm$^{-3}$) attosecond electron-positron jets. This acceleration and bunching process is illustrated in the Supplemental Material movie~\cite{suppmat}. The photon spectrum, shown in Fig.~\ref{FigSolid}(f), bears the signature of this acceleration, since the photons with highest energies (here up to $\approx$1~GeV) are emitted along the laser propagation direction. Finally, we observe that the electron and positron populations are eventually separated spatially by the plasma channel fields~\cite{pukh99}, which are attractive for the electrons and repulsive for the positrons.

A similar simulation overview is shown in Fig.~\ref{FigBeam} in the case of the collision with an electron beam. We observe that the laser is intense enough to put a large fraction of the 10~GeV electrons to a complete stop, by a combination of quantum Compton scattering and Lorentz force. The stopped beam electrons and the electron-positron pairs that they generate are then subsequently accelerated and bunched as before by the Doppler-boosted laser field into relativistic jets. The formation of these dense jets thus appears to be a generic feature of the Doppler-boosted laser-matter interactions close to the Schwinger limit. The Breit-Wheeler electron and positron populations are again partially spatially separated, this time because of the asymmetric temporal profile of the boosted laser (see Supplemental Material~\cite{suppmat}), which tends to push positrons in the positive x direction (see Fig.~\ref{FigBeam}(e)) and electrons in the negative x direction (see Fig.~\ref{FigBeam}(d)).

The spectra displayed in Fig.~\ref{FigBeam}(f) and (g) reveal that the particles can be separated into two populations. The first one is made of particles that travel along the initial electron beam direction. These particles can have very high energies - up to a 10~GeV cutoff corresponding to the initial electron energy - and correspond to particles that have not been stopped by the laser. The second population is made of particles traveling along the laser propagation direction. These particles correspond to charged particles accelerated by the laser and photons that they emit. While the charged particles can be accelerated by the laser to multi-GeV energies, the $\gamma$ photons of this population have rather low energies. This is because ultra-relativisic leptons co-propagating with the laser have a low quantum $\chi$ parameter, which prevents the emission of very high energy photons along the laser propagation direction. The relative importance between these two populations changes strongly with the laser intensity, with more and more particles being stopped and reaccelerated as the laser gets stronger.

\begin{figure}
\includegraphics[width=1.\columnwidth]{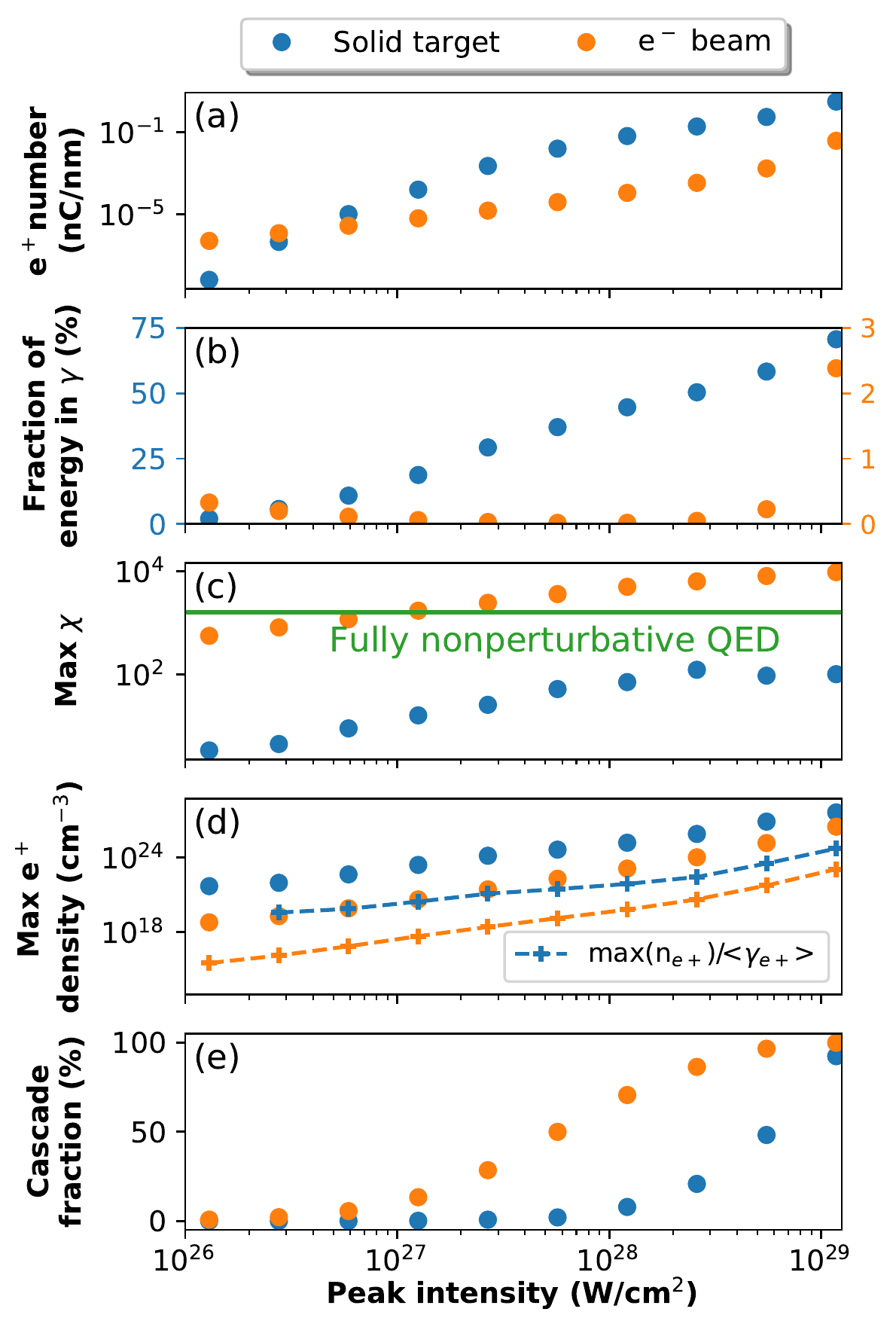}
\caption{Variation of a few selected quantities as a function of boosted laser intensity for solid target (blue) and counterpropagative electron beam (orange) simulations. (a) Total number of generated pairs. (b) Fraction of total initial energy converted into high-energy photons. (c) Highest quantum parameter $\chi$ reached by a target (solid or beam) electron. The threshold for reaching the fully nonperturbative regime is shown in green. (d) Maximum positron density. The crosses show the maximum density divided by the average Lorentz factor of the positrons in the density peak. (e) Fraction of Breit-Wheeler pairs originating from another Breit-Wheeler electron or positron (through the intermediate emission of a high-energy photon).}
\label{FigScan}
\end{figure}

Figure~\ref{FigScan} provides a more quantitative view of the simulation results, for all considered intensities. Panels (a) and (b) are good indicators of the prevalence of SF-QED phenomena during the interaction. We find that using a solid target is more advantageous for maximizing the amount of SF-QED processes, simply because the laser interacts in this case with a much larger number of particles. For intensities exceeding 10$^{27}$~W/cm$^2$, the total number of generated pairs with a solid target is very high (typically in the nC range) despite the small Doppler-boosted laser transverse spot size at focus (approximately 40~nm). We also find that a very significant fraction of the laser energy is converted into high-energy photons (exceeding 50$\%$ for the highest intensity simulations). We are thus entering the so-called QED plasma regime~\cite{zhan20}, for which SF-QED effects are self-consistently coupled with the classical laser-plasma interaction dynamics. In particular, we observe that quantum emission of radiation can in some cases be the dominant laser absorption mechanism, which may eventually open the way towards advanced sources of $\gamma$ photons.

While using a solid target maximizes the amount of SF-QED events, the main advantage of the collision with an electron beam is apparent in Fig.~\ref{FigScan}(c), which shows the maximum quantum parameter reached in the simulation. We acknowledge that the maximum $\chi$ value might not be the most appropriate quantity for characterizing potential deviations to the first-order perturbative dressed loop expansion, but we use it nonetheless because it is easy to define and allows for simple comparisons in a large range of scenarios. We observe that the threshold for entering the fully nonperturbative regime of SF-QED can be largely exceeded during the collision of a boosted laser with a counterpropagative electron beam. These simulations thus demonstrate that the radiation losses in the rising edge of the laser~\cite{krav13,blac19,baum19} can be dealt with so that they do not prevent the observation of the nonperturbative regime, even with a multiple cycle laser. In the simulations, the threshold is reached for intensities around 10$^{27}$~W/cm$^2$. It should be kept in mind however that a more modest intensity of 10$^{25}$~W/cm$^2$, which could be achieved using the focusing naturally provided by the plasma mirror curvature~\cite{vinc19}, would be enough to cross this threshold if a larger scale 100~GeV class conventional accelerator were used to generate the lepton beam, or alternately to reach $\chi$ values exceeding 100 and study the first deviations to the first-order expansion~\cite{dipi20} if combined with a 10~GeV electron beam. In this light, the Doppler boosting technique clearly appears as a very promising solution for probing the fully nonperturbative regime of SF-QED using existing laser technology, even without achieving a diffraction-limited focusing.

A common feature of the performed simulations is the formation of dense attosecond relativistic pair jets accelerated by the laser field itself. The peak density of these jets is plotted in Fig.~\ref{FigScan}(d). We find that the positron density is higher for the solid target interaction, simply because the amount of generated pairs is larger (see Fig.~\ref{FigScan}(a)). The pair jet density reaches in fact enormous values, up to 4$\times$10$^{27}$~cm$^{-3}$ in the simulation with highest intensity, which is 5000 times the number density of the solid target itself. Such values are incomparably higher than those obtained experimentally by irradiating high-Z target, which are at most on the order of 10$^{16}$cm$^{-3}$~\cite{lian15,sarr15}, and even orders of magnitude higher than those obtained in simulations using 100-PW class lasers without Doppler boosting~\cite{neru11,gris16}, which typically scale as $a_0 n_c$, where $n_c$ and $a_0$ are respectively the laser critical density and normalized peak magnetic vector potential.

The observed relativistic pair jets are similar in nature to those existing in the environment of neutron stars and black holes~\cite{chen23}. It is presumed that the interplay between pair plasma effects and SF-QED strongly impacts the radiation emitted by these objects. Reproducing such jets in a laboratory could therefore prove valuable for understanding the signatures of extreme astrophysical objects, especially if pair plasma phenomena occur inside the produced jets. A first criterion to assess whether such effects may occur is to compare the plasma period to the lifetime of the pair jets. In our simulations, the pairs typically reach their maximum density during only 1~fs, which allows us to rewrite this criterion~\cite{chen23} as $n_{e+}/\gamma_{e+} \gg 1.5 \times 10^{20}$~cm$^{-3}$. We can see in the crossed marks of Fig.~\ref{FigScan}(d) that this criterion is satisfied in the solid target simulations for intensities above 10$^{27}$~W/cm$^2$ and in electron beam collision simulations for intensities above 2$\times$10$^{28}$~W/cm$^2$. This is a good first indication that pair plasma phenomena may in fact occur in these jets.

In order to get more physical insights into the simulations, we plot in Fig.~\ref{FigScan}(e) the fraction of "cascade" pairs, defined as the fraction of pairs that originate from a Breit-Wheeler electron or positron (as opposed to pairs originating from a plasma electron). A first observation is that, in the solid target cases the cascade fraction remains negligible for intensities below 10$^{28}$~W/cm$^2$. This means, interestingly, that cascades are not necessary to obtain pair jets with very high densities. Such densities simply come from the bunching of the generated pairs by the laser's attosecond pulses, as illustrated in the Supplemental Material movie.

As the laser field approaches the Schwinger critical field, the emission of $\gamma$ photons by particles accelerated by and co-propagating with the laser, and the decay of these photons into new pair becomes frequent enough to trigger the formation of avalanche type cascades~\cite{bell08,gono22} that very quickly increase the density of the relativistic jets. At the highest intensities, these cascades even start depleting the energy of the Doppler-boosted laser in the solid target case. In the simulation with the highest considered intensity, we find that 75$\%$ of the energy of the strongest attosecond pulse has been absorbed by the relativistic jet in its tail (including the photons emitted by the jet) before the pulse could even reach the plasma surface. This is another example of retroaction of the SF-QED effects on the laser plasma interaction. At even higher intensities, the avalanche cascades develop so fast that we could not complete the simulations, for both the solid target and electron beam cases, because the pair plasma frequency became too high to be resolved, thus triggering the numerical plasma instability~\cite{bird04}.

On a final note, we remark that cascades appear at lower intensities in the electron beam simulations. This is due to the presence of conceptually different shower type cascades~\cite{miro14,gono22}. Such cascades are driven by the very high energy of the incident electrons, which result in the creation of product particles that also have a high energy and can therefore create new particles themselves. However, the particle energy decreases at each generation of the cascade, which stops as soon as the particle energies becomes too low or the particles exit the laser.

In conclusion, we have explored in this letter the interaction of Doppler boosted light with matter at intensities approaching the Schwinger limit. We have found that the interaction with a solid target leads to a plethora of SF-QED events that influence in return the laser-plasma dynamics, corresponding to so-called QED plasma states. The interaction with a relativistic electron beam is complementary and emerges as a pathway to explore the fully nonperturbative regime of SF-QED with existing laser technology. A common feature of both interaction scenarios is the acceleration and bunching of the generated electron-positron pairs into relativistic jets with unprecedented densities.

The richness of the physics that can be glimpsed in these simulations should act as long-term motivation for performing experiments with Doppler boosted lasers focused as tightly as possible. Such a task will require extensive theoretical and numerical efforts to develop the new numerical tools that will be needed to model these interactions more accurately, dealing with any possible commonly used assumption that might break in these extreme conditions (at the very least radiative corrections should be included since $\chi$ values approach or exceed the nonperturbative threshold). In addition, it will be important to understand, and hopefully mitigate, the potential real-world issues (e.g., laser imperfections and aberrations) that could limit the achievable intensities with Doppler boosted lasers.

\begin{acknowledgments}

This research used the open-source particle-in-cell code WarpX https://github.com/ECP-WarpX/WarpX. Primary WarpX contributors are with LBNL, LLNL, CEA-LIDYL, SLAC, DESY, CERN, and TAE. We acknowledge all WarpX contributors. This research was supported by the Exascale Computing Project (17-SC-20-SC), a joint project of the U.S. Department of Energy's Office of Science and National Nuclear Security Administration, responsible for delivering a capable exascale ecosystem, including software, applications, and hardware technology, to support the nation's exascale computing imperative. This research was supported by the CAMPA collaboration, a project of the U.S. Department of Energy, Office of Science, Office of Advanced Scientific Computing Research and Office of High Energy Physics, Scientific Discovery through Advanced Computing (SciDAC) program. This research used resources of the National Energy Research Scientific Computing Center (NERSC), a U.S. Department of Energy Office of Science User Facility located at Lawrence Berkeley National Laboratory, operated under Contract No. DE-AC02-05CH11231 using NERSC award ASCR-ERCAP0022112. This research used resources of the Oak Ridge Leadership Computing Facility at the Oak Ridge National Laboratory, which is supported by the Office of Science of the U.S. Department of Energy under Contract No. DE-AC05-00OR22725. An award of computer time (Plasm-In-Silico) was provided by the Innovative and Novel Computational Impact on Theory and Experiment (INCITE) program. We also acknowledge the ﬁnancial support of the Cross-Disciplinary Program on Numerical Simulation of CEA, the French Alternative Energies and Atomic Energy Commission. This project has received funding from the European Union's Horizon 2020 research and innovation program under Grant Agreement No. 871072. This research was supported by the French National Research Agency (ANR) T-ERC program (Grant No. ANR-18-ERC2-0002).

\end{acknowledgments}


%

\end{document}